\documentclass[aps,prb,twocolumn,showpacs,floatfix]{revtex4}

\usepackage{amsmath}
\usepackage{amssymb}
\usepackage{graphicx}
\usepackage{bm}
\usepackage{mathrsfs}

\begin{document}

\title{Random phases in Bose-Einstein condensates with higher order nonlinearities}  

\author{Mattias Marklund and Padma K. Shukla}
\affiliation{Department of Physics, Ume{\aa} University, SE--901 87 Ume{\aa},
Sweden} 
\affiliation{Institut f\"ur Theoretische Physik IV, Fakult\"at f\"ur
  Physik und Astronomie,  Ruhr-Universit\"at Bochum, D--44780 Bochum,
  Germany} 

\date{\today}

\begin{abstract}
We present a statistical description of Bose-Einstein condensates with general higher order 
nonlinearities. In particular, we investigate the case of cubic-quintic nonlinearities,
of particular interest for dilute condensates. 
The implication of decoherence for the stability properties of the 
condensate is discussed.       
\end{abstract}
\pacs{03.75.Lm, 05.45.-a, 67.40.Vs, 67.57.De}

\maketitle

The basic concept of macroscopic quantum states, such as Bose--Einstein 
condensates (BECs) \cite{bose} and lately also Fermi condensates \cite{fermi}, 
has caught the interest of the physics community, both due to the 
nature of the concept itself but also since the possibilities to perform new
and exciting experiments was early recognized \cite{bose}. 
BECs are normally described by the Gross-Pitaevskii (GP) equation \cite{gp}, 
in which the cubic 
nonlinearity represents two-body forces between the bosons in the 
condensate. There are numerous works on the theoretical foundations
and implications of \cite{Dalfovo-etal,Legget,Pitaevskii-Stringari}. 
When the scattering length is positive, the possibility of dark 
solitons is given via the GP equation. Such dark solitons are stable towards
perturbations in one dimension, while the multi-dimensional case is more complex.
Such dark solitons have also been found experimentally
\cite{Burger-etal}. In the case of a negative effective scattering length,  
the GP equation admits bright solitons, which are prone to 
collapse in dimensions larger than one. Such bright solitons have also been 
experimentally obtained, both in terms of trains of solitons \cite{Strecker-etal}, 
as well as single solitary structures \cite{Khaykovich-etal}. 
The dynamics of these bright soliton structures has also 
been analyzed both analytically \cite{Khawaj-etal} and
numerically \cite{Salasnich-etal}.

If three-body interactions are taken into account, higher order 
nonlinearities will modify the GP equation.   
We may write this as a  generalized NLSE of the form \cite{bar-man} 

\begin{equation}\label{eq:gnlse}
i \hbar \partial_t\psi + \frac{\hbar^2}{2m}\nabla^2 \psi + \alpha |\psi|^2 \psi+
\beta |\psi|^4\psi =0,
\end{equation}
where $\alpha$ and $\beta$ in general is complex-valued, and
$\psi$ is the condensate wave function. 
As $\beta$ goes to zero, we regain the GP equation. 
The real parts of $\alpha$ and $\beta$ corresponds to elastic collisions within the 
condensates, while the imaginary part appears due to inelastic
scattering \cite{Abdullaev-etal}.
In what follows we will neglect the collisional losses, and assume that
$\alpha$ and $\beta$ is real, and can take on both positive and negative 
values. In fact, the coefficient $\alpha$ is proportional to $a$, the  
scattering length, which can be tuned to take on both negative and positive 
values (see, e.g.\ Refs.\ \onlinecite{Gammal-etal} and \onlinecite{Kohler}), 
while $\beta$ is
proportional to $a^4$ for dilute systems \cite{Bedaque-etal,Braaten-etal,Kohler}. 
Thus, in BECs where $a$ is large, the quintic contribution to Eq.\ 
(\ref{eq:gnlse}) may become significant.  Equation
(\ref{eq:gnlse}) can be demonstrated to have solitary solutions in one and two 
dimensions \cite{bar-man}, and appears not only in the physics of BECs, but 
also in, e.g.\ nonlinear optics \cite{zak-etal}. Furthermore, 
Eq.\ (\ref{eq:gnlse}) is a special case of the equation
\begin{equation}\label{eq:gnlse-gen}
  i \hbar \partial_t\psi + \frac{\hbar^2}{2m}\nabla^2 \psi + U(|\psi|^2)\psi =0,
\end{equation}
where $U$ in general is a complex-valued function of the norm
of the wave function squared. 

The stability of solutions to Eq.\ (\ref{eq:gnlse}) towards coherent 
perturbations was analyzed by \textcite{shu-yu}, where the 
growth rate for the modulational instability was found. This growth 
rate signifies the onset of purely growing perturbations, and is thus an important
indication of the possibility of solitary solutions. However, the effects of incoherence, 
e.g.\ a random phase in the wave function, may significantly alter the 
modulational instability and therefore also the onset of inhomogeneity growth,
and is an important issue (for a discussion, see, e.g.\ Refs.\
\onlinecite{Lewenstein-You} and \onlinecite{Pitaevskii}).  
A very direct approach in analyzing the effects of partial coherence lies in
the Wigner formalism \cite{Wigner,Mendonca}. This approach has found uses in 
the study of surface gravity waves in fluids \cite{alber} and electromagnetic waves
in nonlinear media \cite{renato}, in quantum statistical mechanics \cite{kad-baym}, 
in nonlinear optics \cite{Hall-etal}, and in quantum plasmas \cite{Anderson-etal}. 
Here we will apply the Wigner formalism to the problem of partial coherence in the 
modulational instability of higher order nonlinear BECs.
 
In order to analyze the statistical properties of Eq.\ (\ref{eq:gnlse}), 
we may introduce the Fourier transform of the two-point correlation function of the wave 
function, i.e.\ the Wigner function, according to

\begin{equation}\label{eq:wignerfunc}
  F(t,\mathbf{r},\mathbf{p}) = \frac{1}{(2\pi\hbar)^s}\int\,d\bm{\xi}\,
    e^{i\mathbf{p}\cdot\bm{\xi}/\hbar} \langle \psi^*(\mathbf{r} + \bm{\xi}/2,t) 
    \psi(\mathbf{r} - \bm{\xi}/2,t) \rangle
\end{equation}
for the wave function $\psi$, where $s$ denotes the dimensionality of the
problem at hand, the asterisk is the complex conjugate operation, 
and the angular bracket denotes the ensemble average. 
The Wigner function corresponds to a generalized distribution
function for the bosons, and by applying the time derivative to 
Eq.\ (\ref{eq:wignerfunc}) and using  Eq.\ (\ref{eq:gnlse-gen}),
one finds the Vlasov-like equation

\begin{equation}\label{eq:kinetic-gen}
  \partial_tF + \frac{1}{m}\mathbf{p}\cdot\nabla F 
  + \frac{2}{\hbar}U(|\psi|^2)\sin\left(\frac{\hbar}{2}
   \stackrel{\leftarrow}{\nabla}\cdot\stackrel{\rightarrow}{\nabla}_p\right) F = 0 .
\end{equation}
where the $\sin$-operator is defined in terms of its Taylor expansion and
the arrows denote the direction of operation. 
In the case of of a cubic-quintic nonlinearity, such as in Eq.\ (\ref{eq:gnlse}),
we obtain

\begin{equation}\label{eq:kinetic}
  \partial_tF + \frac{1}{m}\mathbf{p}\cdot\nabla F 
  + \frac{2}{\hbar}(\alpha|\psi|^2 + \beta|\psi|^4)\sin\left(\frac{\hbar}{2}
   \stackrel{\leftarrow}{\nabla}\cdot\stackrel{\rightarrow}{\nabla}_p\right) F = 0
\end{equation}
Moreover, the modulus square of the wave function is given by

\begin{equation}\label{eq:mod}
  |\psi|^2 = \int\,d\mathbf{p}\,F(t,\mathbf{r},\mathbf{p}) .
\end{equation}

For the sake of clarity, we now focus on the one-dimensional case.
The stability of equation (\ref{eq:kinetic}) can be analyzed using 
a linearization procedure. Letting $F = F_0(p) + f(p)\exp(ikz - i\omega t)$, where
$|f| \ll F_0$, we linearize Eq.\ (\ref{eq:kinetic-gen}) in order to obtain
the nonlinear dispersion relation

\begin{equation}\label{eq:dispersion-gen}
  1 = -\frac{m}{\hbar k}\frac{dU}{d|\psi_0|^2}\int\,dp\,\frac{F_0(p + \hbar k/2) 
    - F_0(p - \hbar k/2)}{p - \omega m/k} .
\end{equation}
Equation (\ref{eq:dispersion-gen}) is the general dispersion relation for 
matter waves taking into account higher order nonlinearities. 

For the case of a cubic-quintic nonlinearity, Eq.\ (\ref{eq:dispersion-gen})
reduces to
\begin{equation}\label{eq:dispersion}
  1 = -\frac{m}{\hbar k}(\alpha + 2\beta|\psi_0|^2)\int\,dp\,\frac{F_0(p + \hbar k/2) 
    - F_0(p - \hbar k/2)}{p - \omega m/k} .
\end{equation}

For the monochromatic wave case, i.e.\ $F_0(p) = |\psi_0|^2\delta(p - p_0)$, 
we obtain the dispersion relation\cite{shu-yu}

\begin{equation}\label{eq:mono}
  \omega = \frac{p_0k}{m} \pm \left[ \frac{\hbar^2k^4}{4m^2} 
    - \frac{k^2|\psi_0|^2}{m}(\alpha + 2\beta|\psi_0|^2) \right]^{1/2}
\end{equation}
from Eq.\ (\ref{eq:dispersion}). 
Setting $\beta = 0$ in the dispersion relation (\ref{eq:mono}), we obtain the
Bogolubov expression \cite{Bogolubov} for the elementary excitations of the
BEC. The standard method employed in obtaining the result (\ref{eq:mono})
with $\beta = 0$ is to set $\psi = \psi_0 + \psi_1$, where $|\psi_1| \ll |\psi_0|$
and $\psi_0$ is the background state,
and linearizing Eq.\ (\ref{eq:gnlse}), after which the equation may be split
into its real and imaginary part and harmonically decomposed. We note that
the Wigner approach presented in this paper is equivalent in to the 
Bogolubov method in the monochromatic limit.    
Letting $\omega = p_0k/m + i\gamma$ in the
above equation, we obtain the modulational instability growth rate \cite{shu-yu}

\begin{equation}\label{eq:growth}
  \gamma = \left[\frac{k^2|\psi_0|^2}{m}(\alpha + 2\beta|\psi_0|^2) 
    -  \frac{\hbar^2k^4}{4m^2} \right]^{1/2}.
\end{equation}

If the waves are not exactly monochromatic, but have a spectral broadening
due to, e.g.\ a random phase in the background wave function, we may model 
equilibrium condensate spectrum by a Lorentzian distribution \cite{Loudon}

\begin{equation}\label{eq:Lorentz}
  F_0(p) = \frac{|\psi_0|^2}{\pi}\frac{p_T}{(p - p_0)^2 + p_T^2} ,
\end{equation}
where $p_T$ denotes the width of the distribution. The Lorentzian distribution 
solves Eq. (\ref{eq:kinetic}), as well as the more general (\ref{eq:kinetic-gen}), 
and is thus a valid perturbation background. We note that the phase fluctuations
may stem from a variety of perturbations, e.g.\ thermal effects or
quantum fluctuations \cite{Pitaevskii}. The dispersion
relation (\ref{eq:dispersion-gen}) for this case then turns out to be

\begin{equation}\label{eq:dispersion2}
  \omega = \frac{p_0k}{m} \pm \left[ \frac{\hbar^2k^4}{4m^2} 
    - \frac{k^2}{m}\frac{dU}{d|\psi_0|^2}|\psi_0|^2 \right]^{1/2} - i\frac{p_Tk}{m} ,
\end{equation}
which gives the growth rate 

\begin{equation}\label{eq:growth2}
  \gamma = \left[\frac{k^2}{m}\frac{dU}{d|\psi_0|^2}|\psi_0|^2 
    -  \frac{\hbar^2k^4}{4m^2} \right]^{1/2} - \frac{p_Tk}{m} .
\end{equation}
The growth rate (\ref{eq:growth2}) is valid for a general nonlinearity. 
We note that if $dU/d|\psi_0|^2 \leq 0$, there is no modulational instability
growth, and the perturbations are damped. Thus, a minimum requirement
for a positive growth rate for an arbitrary nonlinearity in Eq.\ (\ref{eq:gnlse-gen})
is that $dU/d|\psi_0|^2 > 0$.

In the case of a cubic-quintic nonlinearity $U(|\psi|^2) = |\psi|^2(\alpha + \beta|\psi|^2)$,
Eq.\ (\ref{eq:dispersion2}) reduces to
\begin{equation}
  \omega = \frac{p_0k}{m} \pm \left[ \frac{\hbar^2k^4}{4m^2} 
    - \frac{k^2|\psi_0|^2}{m}(\alpha + 2\beta|\psi_0|^2) \right]^{1/2} - i\frac{p_Tk}{m} ,
\end{equation}
which gives a purely growing modulational instability whose growth rate is

\begin{equation}\label{eq:growth3}
  \gamma = \left[\frac{k^2|\psi_0|^2}{m}(\alpha + 2\beta|\psi_0|^2) 
    -  \frac{\hbar^2k^4}{4m^2} \right]^{1/2} - \frac{p_Tk}{m}
\end{equation}

Comparing with Eq.\ (\ref{eq:growth3}) with Eq.\ (\ref{eq:growth}), 
one clearly sees the damping character 
of the spectral broadening term. We note that since $\alpha$ and $\beta$
may be positive or negative, independent of each other, the instability properties 
crucially depends on the nonlinear terms in the expression (\ref{eq:growth2}). 
If $\alpha, \beta < 0$ (the defocusing case), modulational instability growth is 
not possible, and this corresponds to the well-known stability of dark 
solitary solutions. However, if we have $\alpha < 0$ (defocusing cubic 
nonlinearity), while $\beta > 0$ (focusing quintic nonlinearity), we may have a
new instability regions, not present in the Gross--Pitaevskii equation. If
$\alpha > 0$ (focusing cubic nonlinearity) and $\beta < 0$ (defocusing
quintic nonlinearity) we will have a damping in the growth rate due to three-body
interaction. Note that this damping is quite different from the dissipation
due to inelastic three-body scattering, and has more the character of
Landau damping \cite{Pitaevskii}. It is
also clear that the case $\alpha, \beta> 0$ (the focusing case) gives the maximum
instability growth rate. However, even in this case, a very broad spectral distribution
of the BEC, due to e.g.\ thermal noise, may quench the growth rate considerably,
even removing it all together. We would like to stress that the Wigner method
presented in this paper is a very general approach to partial coherence and spectral 
broadening. Thus, the method is appropriate for a variety of spectral distributions,
e.g.\ Gaussians, as well as the Lorentizian (\ref{eq:Lorentz}), but this will be
pursued in future research.

To summarize, we have presented a perturbation analysis of the statistical properties 
of the generalized Gross--Pitaevskii equation (\ref{eq:gnlse}), by using the 
Wigner formalism. In the case of a random phase of the background wave function
of the condensate, we find that the spectral broadening gives rise to a reduced 
growth rate, as compared to the mono-chromatic case.

\end{document}